\def\year{2020}\relax
\documentclass[letterpaper]{article} 
\usepackage{aaai20}  
\usepackage{times}  
\usepackage{helvet} 
\usepackage{courier}  
\usepackage[hyphens]{url}  
\usepackage{graphicx} 
\usepackage{graphicx}  
\usepackage{amssymb}

\title{Deep Long Audio Inpainting}
\author{Ya-Liang Chang$^*$  $\quad$  Kuan-Ying Lee$^*$  $\quad$  Po-Yu Wu  $\quad$  Hung-yi Lee  $\quad$  Winston Hsu \\ \\
National Taiwan University} 
\begin{document}

\maketitle
\begin{abstract}
Long ($>$ 200 ms) audio inpainting, to recover a long missing part in an audio segment, could be widely applied to audio editing tasks and transmission loss recovery. It is a very challenging problem due to the high dimensional, complex and non-correlated audio features.
While deep learning models have made tremendous progress in image and video inpainting, audio inpainting did not attract much attention.
In this work, we take a pioneering step, exploring the possibility of adapting deep learning frameworks from various domains inclusive of audio synthesis and image inpainting for audio inpainting.
Also, as the first to systematically analyze factors affecting audio inpainting performance, we explore how factors ranging from mask size, receptive field and audio representation could affect the performance.
We also set up a benchmark for long audio inpainting. The code will be available on GitHub upon accepted.
\end{abstract}

\section{Introduction}
Audio inpainting is of significant importance in a broad range of applications to fill in audio gaps of different scales.

Gaps of several to hundreds milliseconds often take place during transmission where packets are subject to frequent events of loss due to unreliable communication channel.
Lots of research has been dedicated to packet loss during transmission and had success tackling gaps at the scale of milliseconds.
For small rates of lost data, sparsity-based \cite{adler2011audio,siedenburg2013audio} sinusodial-based \cite{lagrange2005long}, and autoregressive \cite{oudre2018interpolation} methods are proposed. And for situations with high packet loss rates in speech, \cite{bahat2015self} proposed using an example-based method that exploits prior information from the same user to fill in the gaps. 

\begin{figure}[!htp]
    \centering
    \includegraphics[width=\linewidth]{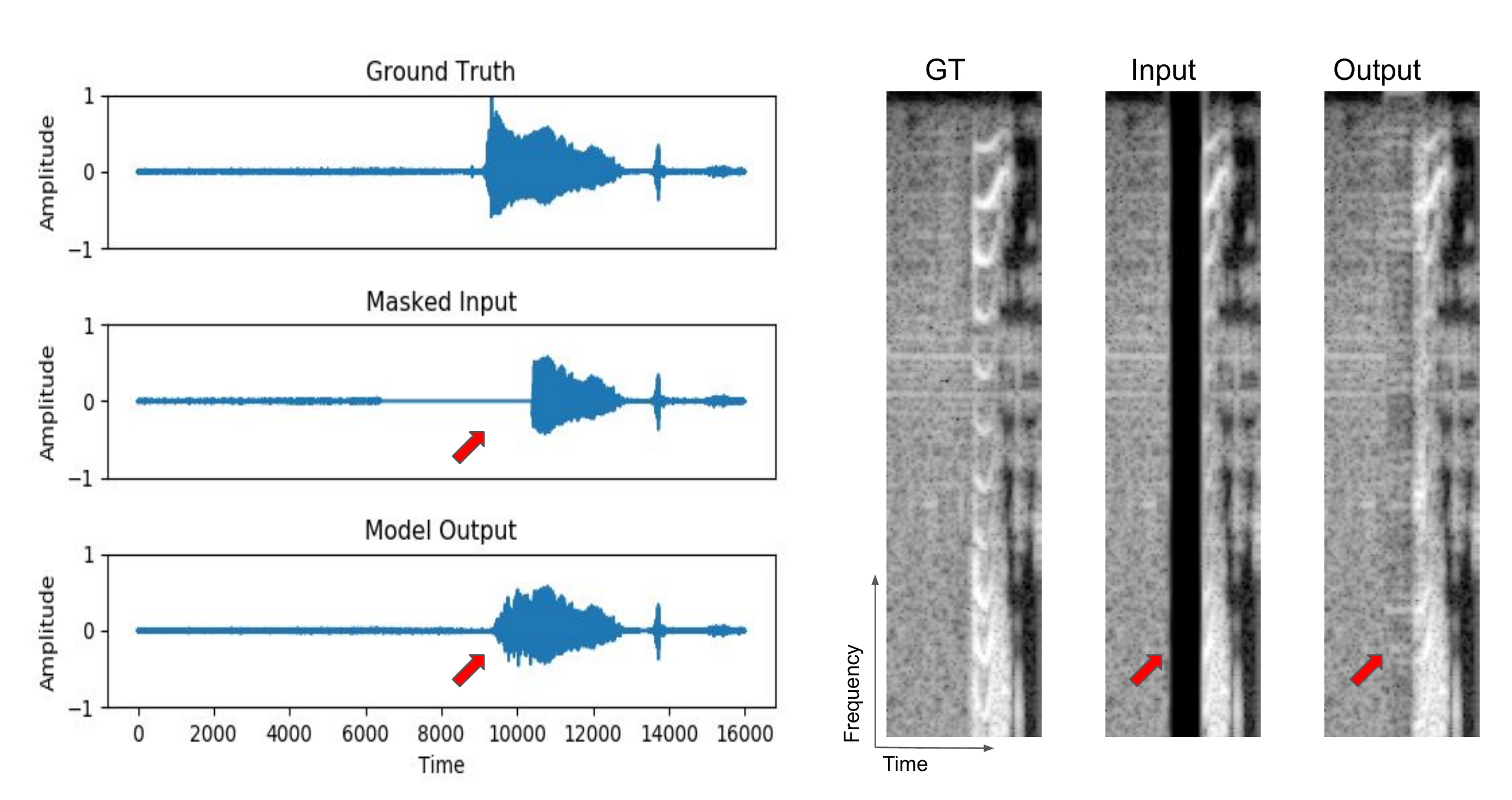}
    \caption{Illustration of the long audio inpainting problem. Given a sound clip with part of it being masked out ($>$ 200 ms), the goal is to recover the masked part. Audio inpainting could be done on either raw waveform (left) or spectrogram (right). Long audio inpainting could be widely used for sound editing tasks such as swear words removal, music editing, etc.}
    \label{fig:teaser}
\end{figure}

Larger gaps spanning for seconds could happen in various applications and cases, such as in music enhancement and restoration. \cite{perraudin2018inpainting} identifies the rather unrealistic assumption often made during shorter-range inpainting that the signal is stationary, which tends not to hold for longer gaps. They harness a similarity graph to obtain similarities between segments and enable second-scale gap filling by substituting the most suitable segment for the gap.

Though these methods have shown quite successes at multiple gap scales, to the best of our knowledge, none have tailored for audio editing, where user could mask out an unwanted segment of an audio, expecting the restoration to sound natural and meaningful (in cases of speech).
While audio editing could be utilized to a broad range of applications, such as removal of environmental noises in a speech or removal of human sound during bird sound recordings, we show current algorithms that targets at second-scale gaps, such as \cite{perraudin2018inpainting} fails when applied onto the scenario\footnote{We do not compare with methods for packet loss since the scale difference is too large.} (cf. Table \ref{tab:benchmark}).

Long ($>$ 200 ms) audio inpainting for editing is a very challenging task.
Firstly, gaps are commonly at the scale of seconds, rendering the algorithms for shorter gaps in vain.
Secondly, in cases of speech, signals are mostly aperiodic and thereby invalidating example based methods such as \cite{perraudin2018inpainting}.
%
%
Thirdly, data are in high dimension ($>$16k/sec) and the correlation between neighboring samples is rather low and thus directly applying state-of-the-art models in image or video inpainting tends not to work well.

Also, while image inpainting has been extensively and explored and promising results based on deep learning frameworks have been demonstrated on large mask inpainting, only a few papers \cite{marafioti2018context} have experimented deep learning on long audio inpainting, let alone discussing how different factors of a neural network could affect the inpainting performance.

Hence, in this work, we take a pioneering step toward long audio inpainting for editing purpose and beyond. As the first to explore the problem, we survey and experiment models from various domains such as image inpainting, Deep Image Prior \cite{ulyanov2018deep} and audio synthesis \cite{prenger2019waveglow}.
%
We also propose two novel frameworks for unconstrained audio inpainting, where we systematically probe into how and to what extent various factors such as gap size, audio representation (either in waveform or spectrogram), receptive field and convolution type (dilated and gated convolution) could impact the inpainting performance,
%
%
%
Also, we setup a benchmark for audio inpainting evaluation and hope it could facilitate future research in this domain.
%
%

Our contributions could be summarized as follows:
\begin{itemize}
    \item We setup a benchmark for long audio inpainting and compare different baselines, based on SC09 dataset of human voice and ESC-50 dataset of natural sound.
    \item We survey and evaluate the possibility of adapting models from different domains for audio inpainting.
    \item We designed novel waveform-based and spectrogram-based models for long audio inpainting.
    \item We experimented different components for deep long audio inpainting, including kernel sizes and model layers.
\end{itemize}

\section{Related Work}
\paragraph{\textbf{Image and video inpainting.}}
Inpainting models aim to restore the masked areas in the image/video, which could be widely used in image/video editing, such as object removal \cite{criminisi2003object,chang2019vornet}. The masked areas are usually given, either by human annotation or segmentation models. The masked area could be a bounding box \cite{wang2018videoinp,yu2018generative}, an object \cite{Huang-SigAsia-2016} or in arbitrary shape \cite{yu2018free,liu2018image,chang2019free}. Many approaches have been proposed to address the inpainting problems, such as diffusion-based ones \cite{bertalmio2000image,bertalmio2001navier} and patch-based ones \cite{barnes2009patchmatch,Huang-SigAsia-2016}. In recent years, deep learning methods become dominant approaches for image inpainting \cite{yu2018free,nazeri2019edgeconnect} and video inpainting \cite{kim2019deep,chang2019learnable} due to the ability to recover unseen parts in an image based on learned data distribution during training. As a baseline, we fine-tune state-of-the-art image inpainting model \cite{wang2018image} to recover missing parts on spectrogram for audio inpainting.

Apart from trained deep image inpainting frameworks, Deep Image Prior \cite{ulyanov2018deep} offers a way to utilize the underlying structure in a untrained network for image restoration and demonstrates a promising result. We also considers it as one of our baselines.

\paragraph{\textbf{Audio inpainting.}}
is to fill gaps in audios, which has been extensively explored under different terminologies \cite{smaragdis2009missing,wolfe2005interpolation}. Many work \cite{marafioti2018context,bahat2015self} dedicates to gaps at the scale of several to tens of milliseconds that are due to packet loss in VoIP, clicks and impulsive noises. 
%
%
In these literature, gaps are at the scale of tens of milliseconds. 
%
%

For gaps ranging from hundreds of milliseconds to several seconds, 
\cite{bahat2015self} utilizes the statistics of recordings from the same user to perform inpainting and \cite{perraudin2018inpainting} proposes to use a graph to capture spectral similarity of different segments in the signal, where the most suitable one is used for inpainting.
Nevertheless, \cite{perraudin2018inpainting} is only practical for signals with repeated patterns (e.g. music) and tends to fail on aperiodic signals like speech (cf. Table \ref{tab:benchmark}), while \cite{bahat2015self} could only handle speech with the same identities.
We still set \cite{perraudin2018inpainting} as one of the baselines since \cite{bahat2015self} is not suitable for datasets like ESC-50 \cite{piczak2015dataset} for audios inside are all natural sounds.


While there is also work on speech inpainting \cite{prablanc2016text}, we do not compare with it as it requires text to perform inpainting.

\paragraph{\textbf{Audio synthesis.}} 
is to generate audios either unconditionally or based on given cues.
%
A pioneering work is WaveNet \cite{oord2016wavenet} which achieves longer-range dependency with enlarged receptive fields through dilated convolution. Yet, one drawback for direct generation of audio samples through auto-regressive structures is its low speed.

Hence, many work have since built upon it to improve the generation speed. Still under auto-regressive structure, WaveRNN \cite{kalchbrenner2018efficient} substitutes RNN for the stack of convolutions in \cite{oord2016wavenet}. Another prevalent approach is to generate an intermediate spectrogram before converting it to the final audio \cite{prenger2019waveglow,donahue2018adversarial}.

\begin{figure*}[!htp]
    \centering
    \small
    \includegraphics[width=\linewidth]{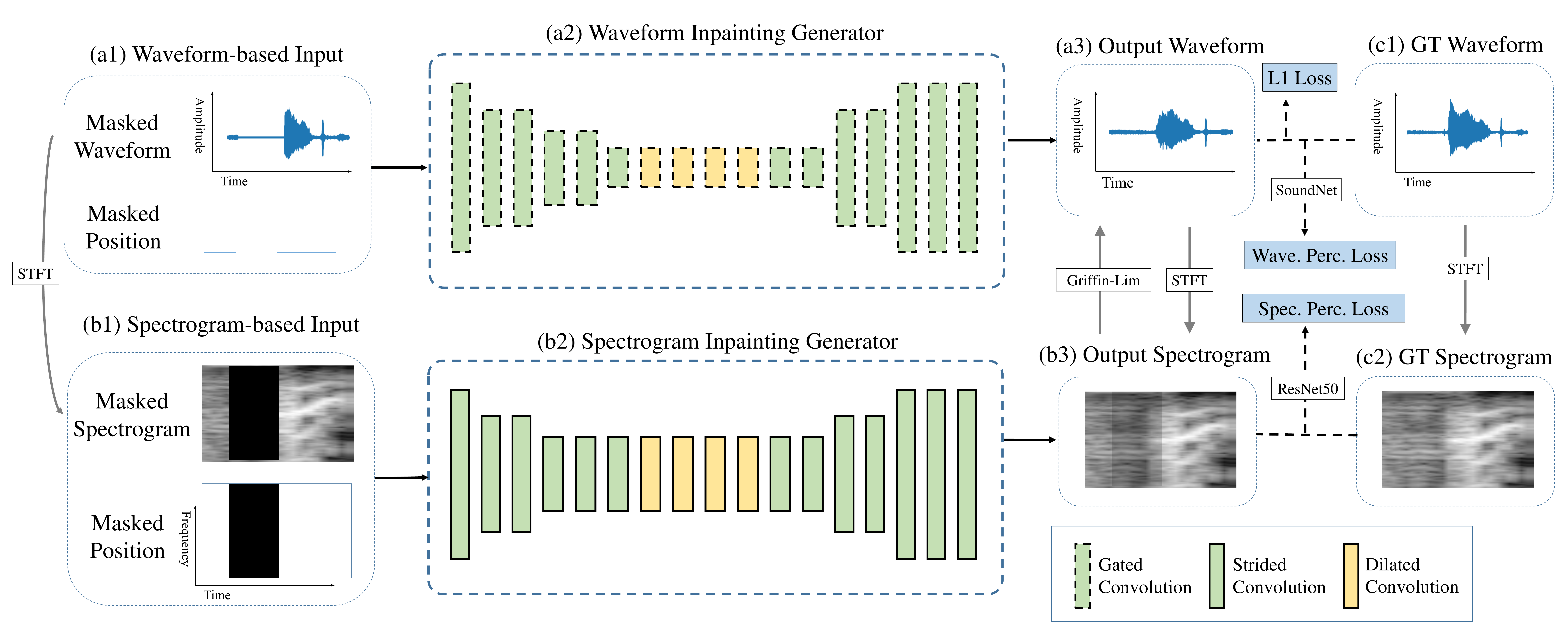}
    \caption{Overall model architecture. (a) Waveform inpainting model takes masked raw waveform and mask as input and directly generates inpainted waveform. (b) Spectrogram inpainting model first transforms masked waveform to spectrogram, inpaints it as an image, and then transform it back to waveform with the Griffin-Lim algorithm. L1 loss is calculated between (a3) and (c1), while perceptual losses are derived  using pre-trained models (SoundNet/ResNet50) to extract features for waveform (a3, c1) / spectrogram (b3, c2). Note that we experiment different design of (a2/b2) in the ablation study.}
    \label{fig:model_architecture}
\end{figure*}

Though the goal of audio synthesis is different than that of audio inpainting, they could both generate audios conditionally.
Hence, we consider Waveglow \cite{prenger2019waveglow} as one of the baselines and train the vocoder to generate inpainted audio given a masked spectrogram instead of a complete one.

\section{Proposed Method}
\subsection{Definition}
For audio inpainting, we take an input audio sequence $\{A_t  \mid  t=1 \dots n\}$ with a mask $\{M_t  \mid  t=i \dots j, 1<i<j<n\}$ as input. The masked samples are set to be zeros. The model will recover the masked samples and generate the output audio $\{O_t  \mid  t=1 \dots n\}$, and the goal is to minimize the loss between $O_t$ and $A_t$.

\subsection{Spectrogram Inpainting Model}
For our spectrogram inpainting models, we first transform each the $A_t$ into a spectrogram $S_t$ by short-time Fourier transform (STFT) with window width $\omega$, treat it as a special image and thereby considering audio inpainting problem as a special image inpainting problem to recover the missing parts in the spectrogram. Then the recovered spectrogram will be transformed back to waveform as $O_t$ by 
Griffin-Lim algorithm \cite{griffin1984signal} for comparison.

The spectrogram inpainting model architecture is based on state-of-the-art image inpainting model \cite{yu2018free} (see Fig. \ref{fig:model_architecture} (b2)). However, unlike natural images where x and y dimensions have similar scale and meaning, the time and frequency dimension in spectrograms have a significant difference. Therefore, we explore different components to deal with convolutions on spectrogram (see Fig. \ref{fig:model_architecture}).

\subsection{Waveform Inpainting Model}
Our waveform inpainting models directly takes masked raw waveform as input and generate recovered waveforms as outputs. However, different from spectrograms, the raw waveforms are in much higher dimension ($>$16k/sec). If we have a one-second audio clip at a sample rate of 16 kHz, over 61 samples are needed to capture a single cycle of the 261.63 Hz sinusoid, C4 of the musical note. As discussed in the previous works \cite{aytar2016soundnet,oord2016wavenet,donahue2018adversarial}, larger convolutional kernels and strided/dilated convolutions are often needed to deal with audio signals as they could increase the receptive field. On the proposed waveform inpainting model, we thus experiment with gated/dilated convolution, 

\paragraph{\textbf{Gated convolutions.}}
For each convolutional layer in waveform-based models (Fig. \ref{fig:model_architecture} (a2)), we adopt gated convolution \cite{yu2018free} to softly attend on the masked areas:
\begin{equation}
Output = \sigma(W_g * x)  \phi(W_f * x)
\end{equation}
where $x$ is the input feature, $W_g$ is the gating kernel, $W_f$ is the feature kernel, $sigma$ is the sigmoid function to restrict the soft gating values between 0 (invalid) and 1 (valid), $\phi$ is the activation function (LeakyReLU), and $*$ is the convolution operation.
Note that a similar idea, gated activation \cite{van2016conditional} is also found useful for the audio generation task such as WaveNet \cite{oord2016wavenet}.

\subsection{Loss Functions} \label{sec:loss_functions}
\paragraph{\textbf{Masked $l_1$ loss ($Ml_1$).}}
The $l_1$ loss focuses on low-level features and is widely used for both image and video inpainting models \cite{liu2018image,wang2018videoinp,chang2019free}. We apply the $l_1$ loss on the masked area:
\begin{equation} \label{l1_loss}
L_{Ml_1} = \mathop{\mathbb{E}_{t}}[ M_{t} |O_{t} - A_{t}|]
\end{equation}

\paragraph{\textbf{Perceptual loss on waveforms.}} $l_1$ loss often leads to blurry results \cite{yu2018free,chang2019free}, so we adopt the perceptual loss \cite{gatys2015neural} originally used for style transfer to enhance the audio quality. It is also used for image inpainting \cite{liu2018image,yu2018free}, video inpainting \cite{chang2019free} and super-resolution \cite{johnson2016perceptual,ledig2017photo}.

Similar as pre-trained VGG \cite{simonyan2014very} on ImageNet \cite{russakovsky2015imagenet} for image perceptual loss, we use pre-trained SoundNet and fine-tune it on our benchmark dataset with classification task for audio perceptual loss:
\begin{equation} \label{percputal_loss}
L_{perc} = \sum_{t=1}^{n} \frac{|\Psi^{O_{t}} - \Psi^{V_{t}}|}{N_{\Psi^{V_{t}}}}
\end{equation}
where $\Psi$ is the features extracted from last layer before fully-connected of the fine-tuned SoundNet. Note that we follow a similar fashion as how \cite{chen2018visually} uses a pretrained SoundNet to compute audio perceptual loss.

\paragraph{\textbf{Perceptual loss on spectrograms.}}
Aside from waveforms, we propose to consider perceptual loss on spectrograms, as image classification is relatively easier to learn. We transform waveforms to spectrograms with STFT, which are then used to fine-tune a  ResNet50 \cite{he2016deep} pretrained on ImageNet \cite{russakovsky2015imagenet} for audio classification. The fine-tuned ResNet50 then serves as a feature extractor for perceptual loss on spectrograms, as shown in Equation \ref{percputal_loss}.

\section{Experimental Results}
\subsection{Datasets}
\paragraph{\textbf{SC09.}} SC09 dataset is a subset of the Speech Commands Dataset \cite{warden2018speech} that contains single spoken word from zero to nine by different speakers in uncontrolled environments.
Since its first proposal by \cite{donahue2018adversarial}, it has been used in many audio generation research \cite{donahue2018adversarial,marafioti2019adversarial} and often regarded as the most common baseline in the area. (just as MNIST dataset \cite{lecun1998mnist} in written digit recognition, although examples in SC09 are more complicated ($\mathbb{R}^{16000}$) than MNIST ($\mathbb{R}^{28*28=784}$))

\paragraph{\textbf{ESC-50.}} ESC-50 dataset \cite{piczak2015dataset} is a labeled dataset for environmental sound classification, including 2000 5-second long environmental audio recordings of 50 semantic classes (40 examples per class) from 5 categories: animals, natural soundscapes \& water sounds, human non-speech sounds,	interior/domestic sounds and exterior/urban noises.
Compared to SC09, examples in ESC-50 are more repetitive and thus easier for patch- and example-based methods but harder for learning-based methods for it has more classes and fewer data per class.

\subsection{Benchmark Procedure}
We setup the long audio inpainting benchmark to compare baselines, including WaveGlow \cite{prenger2019waveglow}, SimilarityGraph \cite{perraudin2018inpainting}, DeepPrior \cite{ulyanov2018deep} and GMCNN \cite{wang2018image}.
For SC09, we train all the models on the whole training set with random masking of 0.2 second and without any data augmentation. We perform evaluation on the testing set with fixed mask from 0.4 $\sim$ 0.6 second.
For the ESC-50 dataset, we train models with the first 1600 sound clips (first to fourth fold) with random masks of 0.4 second.
Sound clips are copied twice to 10 seconds and then randomly cropped to 5 seconds during training. The first 200 sound clips of the fifth fold is used for validation while testing is done on the other 200 sound clips with fixed mask from 3.0 $\sim$ 3.4 second. For models that require longer inputs, we apply zero padding.
Note that after inpainting, we paste the unmasked segments from input to the output.

\subsection{Baseline Implementation Details}
\subsubsection{WaveGlow}
is a flow-based vocoder that transforms a melspectrogram to its corresponding final waveform. It combines essence of WaveNet and Glow and directly learns the data distribution. We modify it to take a masked melspectrogram instead of a complete one as input, using the codes provided by NVidia\footnote{\url{https://github.com/NVIDIA/waveglow}}.
We train each model for 100000 epochs with batch size 2 and 4 for SC09 and ESC-50 respectively.

\subsubsection{Deep Image Prior}
performs well on several image restoration tasks including inpainting by simply using the structure of a neural network and the corrupted image without any training beforehand.
We harness the inpainting script in Github\footnote{\url{https://github.com/DmitryUlyanov/deep-image-prior}} to inpaint the masked spectrogram. We change the target iteration from 6001 to 4001 to reduce the processing time while still maintaining the quality of the audio.

\subsubsection{GMCNN}
is one of the state-of-the-art image inpainting framework that features a multi-column neural network that could model different image components and extract multi-level features to aid inpainting.

We experiment with the framework provided in Github \footnote{\url{https://github.com/shepnerd/inpainting_gmcnn}} and  modify the model to take a masked spectrogram as input instead of a 256 * 256 image with RGB channels.
Since the spectrogram is one channel, we firstly modify the pretrained model by changing the first layer of the generator to a conv layer with one input channel, the last decoding layer to a conv layer with one output channel and the first layer of both the global and local discriminator to a conv layer with one input channel. These conv layers are all initialized randomly with normal distribution.
We then finetune the model for 40 epochs using the default settings. 

\subsubsection{SimilarityGraph}
is an example-based framework that targets particularly at long gaps in music. It detects spectro-temporal similarities among unmasked data to the masked area, solving case when adjacent segments fail to provide a solution.

We harness the demo website\footnote{ \url{https://epfl-lts2.github.io/rrp-html/audio_inpainting/}} to perform inpainting. Since their framework requires that the mask to be at least 3 seconds away from the start and the end of the audio, we perform duplication before uploading audio to the website. For SC09, we duplicate both the front and rear 0.4 second for 8 times, generating an audio that is 0.4 * 16 + 0.2 = 6.6 second long (Hence, the mask is from 3.2 to 3.4 second). And for ESC-50, we duplicate only the rear 1.6 second twice, generating an audio that is 3 + 0.4 + 1.6 * 2 = 6.6 second long (Hence, the mask is from 3 t o 3.4 second).
After inpainting, we extract the segment that corresponds to the original audio. Note that since the algorithm replaces the masked part with a audio segment from the same signal, the position of the original audio might shift slightly.

\subsection{Evaluation Metrics}
To evaluate different methods numerically, we calculate the masked $l_1$ error (Eq. \ref{l1_loss}) and perceptual distance (Eq. \ref{percputal_loss}) between the outputs and ground truths on waveforms and sepctrograms, as explained in Sec. \ref{sec:loss_functions}. For fair comparison of perceptual distance on waveforms, we finetune another SoundNet \cite{aytar2016soundnet} and VGG16 \cite{simonyan2014very} for perceptual distance.
Please see Table \ref{tab:perceptual_loss_classification} for detailed settings of different backbones (pre-) trained on SoundNet and ResNet50 for perceptual loss on waveforms, spectrograms respectively.
In addition, we also report the structural similarity (SSIM) index \cite{wang2004image} on spectrograms. The inference time is reported in terms of how many sound clip could be processed per second on Intel(R) Xeon(R) Gold 6154 CPU \@ 3.00GHz with a single V100 GPU.

\begin{table}[!htp]
 \centering
 \small
 \begin{tabular}{|c|c|c|c|c|c|}
 \hline
\begin{tabular}[c]{@{}c@{}}Dataset\end{tabular} &
\begin{tabular}[c]{@{}c@{}}Type\end{tabular} &
\begin{tabular}[c]{@{}c@{}}Model\end{tabular} &
\begin{tabular}[c]{@{}c@{}}Pretr.\end{tabular} &
\begin{tabular}[c]{@{}c@{}}Param. \end{tabular} &
\begin{tabular}[c]{@{}c@{}}Acc. \end{tabular}  \\
\hline \hline 
SC09 & Wave. & SoundNet & \checkmark & 14.3M & 93.4\%\\
&& SoundNet &  $\times$  & 14.3M & 91.0\%\\
&Spec. & VGG16 & \checkmark &  134.3M & 96.0\% \\
&& ResNet50 & \checkmark & 23.5M & 96.3\% \\
&& ResNet50 & $\times$ & 23.5M & 94.7\% \\
\hline 
ESC-50 & Wave. & SoundNet & \checkmark & 14.7M & 66.3\% \\
&& SoundNet &  $\times$  & 14.7M & 61.0\%\\
&Spec. & VGG16 & \checkmark & 134.4M & 83.5\%\\
&& ResNet50 & \checkmark & 23.6M & 82.0\%\\
&& ResNet50 & $\times$ & 23.6M & 77.0\% \\
\hline 
 \end{tabular}
 \caption{Sound classification accuracy for perceptual losses/metrics on SC09 and ESC-50 testing set.}
\label{tab:perceptual_loss_classification}
\end{table}

\begin{figure}[!htp]
    \centering
    \includegraphics[width=0.9\linewidth]{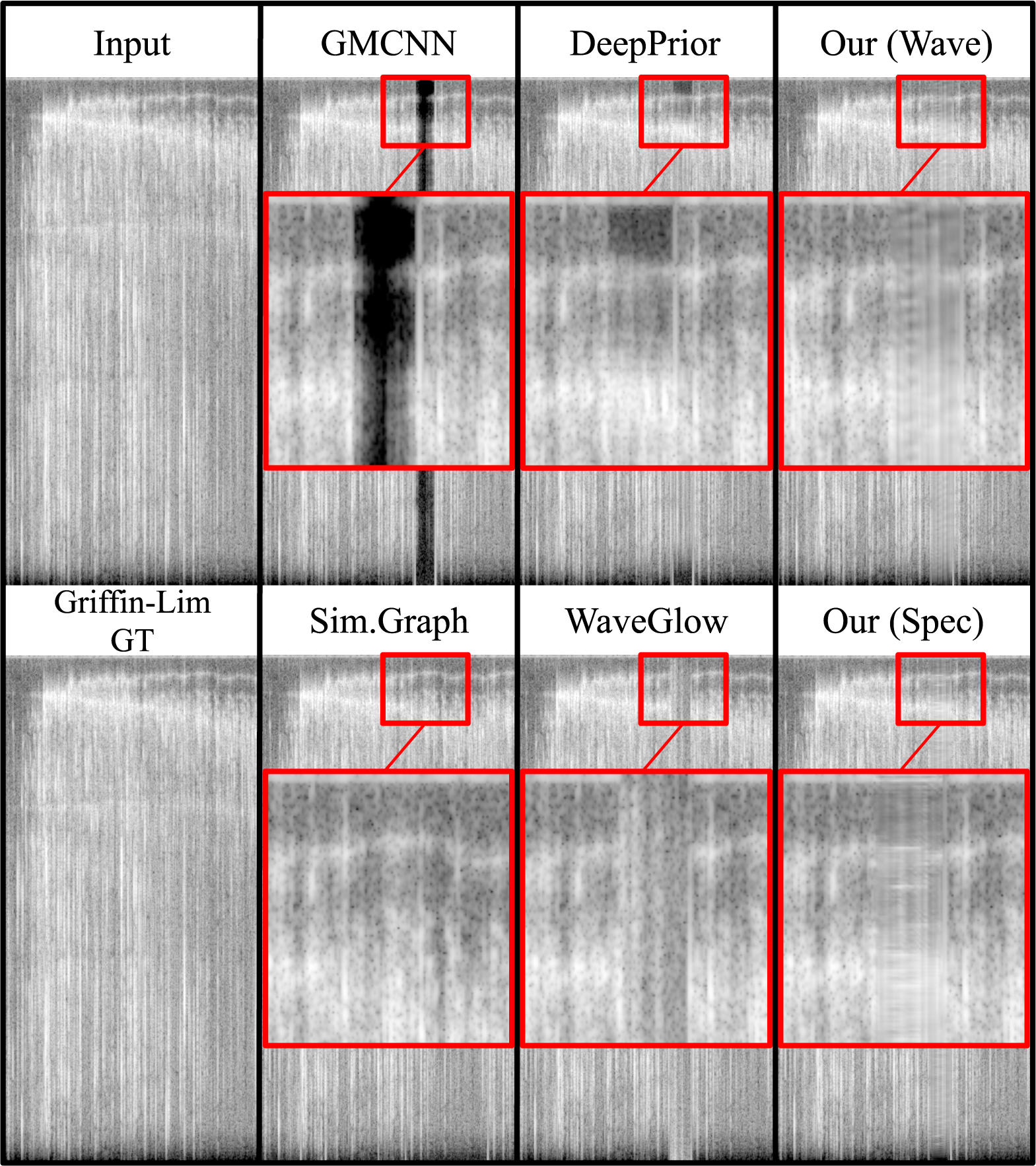}
    \caption{Spectrograms of audio inpainting results.  SimilarityGraph performs well as sounds in ESC-50 tend to embed repetitive structures for example-based method to exploit. Also, we show that despite its long processing time, DeepPrior also excels at spectrogram inpainting aside from image inpainting.
    }
    \label{fig:qualitative}
\end{figure}

\subsection{Quantitative Results}
\paragraph{\textbf{Sound classification.}}
For perceptual losses evaluation, we finetune SoundNet or train ResNet from scratch on audio classification with cross entropy for waveform and spectrogram respectively.
The classification accuracy of each model is reported in Table \ref{tab:perceptual_loss_classification}. We could observe that for ESC-50, spectrogram based classification models outperform those based on waveform, while in SC09, this does not hold.
%


This might be caused by the input dimension. That is, raw waveforms in ESC-50 have more samples (22050 $\times$ 5 = 110250 samples) than those in SC09, making it harder for models based on 1-d convolutions with limited receptive field size to extract high-level features.
On the contrary, after STFT, spectrograms are only in 1024 $\times$ 400 and 1024 $\times$ 80 for ESC-50 and SC09 respectively, which are reasonable sizes for image classification models and the size difference is smaller for the two datasets.

Another interesting point is that the pre-trained weights on ImageNet boost spectrogram classification for about 5\%, even though the applications and domains are quite different (3-channels natural images vs 1-channel spectrograms). It possibly imply that pre-training on other datasets such as Google Audio set \cite{gemmeke2017audio} could further improve the perceptual losses and metrics.

\begin{table*}[!htp]
\centering
\small
\begin{tabular}{|c|c|c|c|c|c|c|c|c|c|}
\hline
& \multicolumn{4}{c|}{SC09}  & \multicolumn{4}{c|}{ESC-50} &  \\ \hline
Method  & ML1 $\downarrow$ & SSIM $\uparrow$& \begin{tabular}[c]{@{}c@{}} Wave. \\ P. Dist.$\downarrow$ \end{tabular} & \begin{tabular}[c]{@{}c@{}}Spec. \\P. Dist. $\downarrow$\end{tabular}  & ML1 $\downarrow$ & SSIM $\uparrow$& \begin{tabular}[c]{@{}c@{}}Wave. \\ P. Dist.$\downarrow$ \end{tabular} & \begin{tabular}[c]{@{}c@{}}Spec. \\P. Dist. $\downarrow$\end{tabular} & \begin{tabular}[c]{@{}c@{}}Infer.\\ Speed\textsuperscript{*} $\uparrow$\end{tabular} \\ \hline \hline
Masked Input & 0.011125 & 0.675040  & 0.006231 & 0.079655 &  0.067510  &   0.648608   & 0.007740 &  0.542866  &  -- \\ \hline
Griffin-Lim GT  & 0.021293  & 0.808092 & 0.004214 & 0.021372  & 0.004539 & 0.978142 & 0.000784 & 0.007324  &  -- \\ \hline \hline
WaveRNN   &  $\times$  &   $\times$   & $\times$   &  $\times$  &  $\times$  & $\times$ & $\times$ & $\times$   & $\times$  \\ \hline
WaveGlow     &  0.013689  &  0.730689 & 0.004494 &  0.077139   &  0.003048  &   0.929394   & 0.000821 & 0.035776 &  2.058\\ \hline
Sim.Graph  &  $\times$   & $\times$  & $\times$  & $\times$  &  0.004478  &  0.697933    & 0.003229 &  0.115829 &  0.039\\ \hline
GMCNN   & \textbf{0.010769} &  0.695439 & 0.004866 & 0.073790 & 0.002738 &  \textbf{0.935945} & 0.000728 & 0.031737 &  63.09 \\ \hline
DeepPrior    &  0.010940 & 0.719634 & 0.004607 & 0.067535 & 0.004175 &    0.951980  & \textbf{0.000499} & \textbf{0.017755} & 0.002\\ \hline \hline
\begin{tabular}[c]{@{}c@{}}Ours (Spec.) \\ (L1)\end{tabular}  & 0.012073 & 0.422832 & 0.008016 &  0.080591  & 0.003148  &  0.727943      &0.002362 & 0.154318 & \textbf{106.38} \\ \hline
\begin{tabular}[c]{@{}c@{}}Ours (Spec.) \\ (L1+SpecP)\end{tabular} & 0.017605 &  0.384210 & 0.006532 & 0.066645 &  0.003159 &   0.721920   & 0.002334  & 0.152135 & \textbf{106.38} \\ \hline
\begin{tabular}[c]{@{}c@{}}Ours (Wave.) \\ (L1)\end{tabular} & 0.010860 &  0.696274 & 0.005817 &  0.071622 &  \textbf{0.002696} &  0.923965   &  0.000888 & 0.035509 & 92.93 \\ \hline
\begin{tabular}[c]{@{}c@{}}Ours (Wave.) \\ (L1+WaveP)\end{tabular} & 0.013796   &  \textbf{0.775181}    & \textbf{0.002909} &  \textbf{0.051784}  &  0.002931  &  0.923112   &  0.000859 & 0.035125 & 92.93 \\ \hline
\end{tabular}
 \caption{Long audio inpainting benchmark results. $\times$: WaveRNN fails to converge; SimilarityGraph algorithm fails to find a solution for most cases in SC09. \textsuperscript{*}The infer. speed is how many SC09 samples an algorithm could process in a minute.}
\label{tab:benchmark}
\end{table*}

\paragraph{\textbf{Long audio inpainting benchmark results.}}
The long audio inpainting benchmark results are presented in Table \ref{tab:benchmark}. We could observe that our models perform reasonably well on both SC09 and ESC-50 for all metrics. Still, we find that all the evaluation metrics could not totally reflect human perception. For example, the STFT + Griffin-Lim process would seriously damage the SSIM score even when the input is simply ground truth (see the Griffin-Lim GT column); the perceptual distances are not affected by the process, but it may not totally reflect the amplitude (see Fig. \ref{fig:qualitative}: GMCNN has low perceptual distance).
On the other hand, although results from SimlarityGraph are quite natural to humans (since the mask is pasted with the other part of the sound clip), its performance is not as good in all metrics as the filled in contents are different.
Surprisingly, image inpainting models GMCNN and DeepPrior outperform other baselines in all metrics (note that DeepPrior does not require training), whereas vocoders like WaveGlow and WaveRNN are not as good. It indicates that general image inpainting models could highly likely be adapted to handle spectrogram as well and our spectrogram-based model still have a large space to improve, such as the kernel size, training loss, etc. 
Also, though the perceptual loss we apply does help a little bit, it generally does not lead to large improvement.

\subsection{Qualitative Results}
We compare output spectrograms from different baselines qualitatively in Fig. \ref{fig:qualitative}. The spectrograms show a sound of water filling a container in five seconds with a 0.4 second mask at three second.
The sound of Griffin-Lim GT has no mask and thus depicts how the spectrogram would look like after undergoing Griffin-Lim algorithm.

In baselines, we discover that SimilarityGraph and DeepPrior perform well on inpainting the water sound. The environmental sounds in ESC-50 contain a lot of repeating structures.
Due to this reason, the result of SimilarityGraph intuitively sounds great by with its copy and paste solution. Note that SimilarityGraph fails on most SC09 cases, as there are no repetitive structures that could be pasted in cases of zero to nine.
With results from DeepPrior, which is good at extrapolating local correlation, we show that sounds, like images, have local property as well.

We found that DeepPrior does surprisingly well on audio inpainting, extrapolating implicit structures embedded in spectrograms. WaveGlow inpaints with sheer noise and GMCNN fails and inpaints sheer silence.

Our proposed method inpaints meaningful elements instead of sheer noise or pure silence in the masked part, as shown in Fig. \ref{fig:qualitative}.
Compared to baselines like DeepPrior and SimilarityGraph, where results are more clear and natural than that our results. Nevertheless, the DeepPrior needs much more inference time than other else and the SimilarityGraph highly constrains on specific tasks due to its copy and paste solution.

\subsection{Ablation Study}
In this section, we experiment with different parameters, inclusive of mask ratio and receptive field to explore how these factors may affect our model. Note that we perform all the following experiments on ESC-50.

We perform two set of experiments. In the first one, we fix the length of the mask and alter the receptive field by configuring the network architecture. And in the second experiment, we fix the receptive field and see how different mask sizes actually impact the performance.

\begin{table}[!htp]
 \centering
 \small
 \begin{tabular}{|c|c|c|c|c|c|}
 \hline
\begin{tabular}[c]{@{}c@{}}Masked\\Time (s)\end{tabular} &
\begin{tabular}[c]{@{}c@{}}Masked\\Field\end{tabular} &
\begin{tabular}[c]{@{}c@{}}Receptive\\Field\end{tabular} &
\begin{tabular}[c]{@{}c@{}}L1\\loss.\end{tabular} &
\begin{tabular}[c]{@{}c@{}}SpecP\\Error\end{tabular} &
\begin{tabular}[c]{@{}c@{}}Suc.\end{tabular} \\
\hline \hline 
0.1 & 40 & 21 & 0.0272 & 0.131 & $\times$ \\
0.1 & 40 & 29 & 0.0236 & 0.114 & \checkmark \\
0.1 & 40 & 45 & 0.0217 & 0.101 & \checkmark \\
0.1 & 40 & 61 & 0.0216 & 0.105 & \checkmark \\
0.1 & 40 & 77 & 0.0209 & 0.108 & \checkmark \\
0.1 & 40 & 93 & 0.0206 & 0.098 & \checkmark \\
0.1 & 40 & 109 & 0.0214 & 0.117 & \checkmark \\
0.1 & 40 & 125 & 0.0218 & 0.101 & \checkmark \\
\hline
0.15 & 60 & 77 & 0.0338 & 0.1571 & \checkmark \\
0.16 & 64 & 77 & 0.0380 & 0.1754 & \checkmark \\
0.17 & 68 & 77 & 0.0446 & 0.2102 & $\times$ \\
0.18 & 72 & 77 & 0.0447 & 0.2003 & $\times$ \\
0.19 & 76 & 77 & 0.0480 & 0.2305 & $\times$ \\
0.2 & 80 & 77 & 0.0552 & 0.2390 & $\times$ \\
0.25 & 100 & 77 & 0.0676 & 0.2855 & $\times$ \\
\hline 
 \end{tabular}
 \caption{L1 loss and Perceptual error of different masked time(sec) and receptive field on ESC-50. The masked field and receptive field are based on time axis. Error is calculated on validation set. Success presents whether the model successfully inpainted the whole masked part or failed on certain field. Fail means no change on the magnitude of inpainting part lasting a period even the model inpainted most of the masked field.}
\label{tab:mask_ratio_perceptive_field}
\end{table}
%
%
%
We evaluate on models that are trained for 50 epoch with L1 loss and Spectrogram perceptual metrics (see Table \ref{tab:mask_ratio_perceptive_field}).

\paragraph{\textbf{Receptive field.}}
In our proposed structure, increasing the depth of the network enlarges the receptive field.
%
According to our experiment results, the receptive field has to be larger than a certain threshold in order to inpaint the whole mask.
Nevertheless, after reaching a certain size, enlarging the receptive field has little benefit or even negative effect for training. That indicates, after some threshold, our model is complicated enough and is able to restore the mask.

\paragraph{\textbf{Mask ratio.}}
We train several models by altering the mask length from 0.1 to 0.25 second and keep the receptive field fixed.
We found that our model could adapt to different mask lengths (from 0.1 to 0.16 seconds), with a fixed receptive field, as long as the mask size is smaller or equals to receptive field. This, on the other side, again confirms that the receptive field has to be at least similar to the mask size to perform successful inpainting.
%
%
Also by observing the failure cases, the inpainted sound at the mask position trailed off at first, vanished at the middle, and then appeared in the end. It indicates that if the mask field is too big, the receptive field will not be sufficient to gather enough information to rebuild the whole masked part.

\section{Discussion and Future Work}
\paragraph{\textbf{Receptive field and model architecture.}}
To the best of our knowledge, we are the first to evaluate different baselines and model architectures for deep long audio inpainting
We also discuss the effect of different mask ratios and receptive field in the ablation study. However, compared to image classification/inpainting, there is very little research working on the model architecture of deep audio tasks.
Current architectures are very diverse, including different stride/dilation/kernel sizes, while ESC-50 is not large and diverse enough as ImageNet for comparison.
Further experiments could be done to find out a common structure for audio perceptual loss and waveform/spectrogram based audio inpainting, possibly through neural architecture search \cite{zoph2016neural}.

\paragraph{\textbf{More general datasets or datasets with other clues.}}
For the proposed benchmark, we compare methods on SC09 and ESC-50, corresponding to complicated short human voices and repetitive natural sounds. Nevertheless, in real-world scenarios, there are many more kind of sounds with longer periods and more complex/simple structures, such as speech and music. Our benchmark currently does not cover enough datasets for general audio editing. Also, in many cases other clues such as texts, images and videos are given at the same time, which could possibly assist long audio inpainting.

\paragraph{\textbf{Spectrograms to waveforms.}}
In this work, we apply the Griffin-Lim algorithm to turn spectrograms back to waveforms as in the audio synthesis \cite{donahue2018adversarial} and text-to-speech \cite{tachibana2018efficiently} task. The reconstructed waveforms are similar to the original ones but with a slight loss (see the Griffin-Lim GT in Table \ref{tab:benchmark}).
It's worth mentioning different from the two tasks, most phases in long audio inpainting are available and could be used for better phase estimation of the masked area to transform spectrograms back to waveforms. The model could learn better to recover the missing phases in the masked area with hints from the surrounding phases and thus better waveform reconstruction.

\paragraph{\textbf{GAN loss.}}
Recently, many image/video inpainting \cite{yu2018free,chang2019free} and audio synthesis works \cite{donahue2018adversarial} adopt the generative adversarial network (GAN) \cite{goodfellow2014generative} to enhance output realness. However, in our experiments, the GAN loss does not help much for our models. How to incorporate GAN and other loss functions to make output sounds more realistic is a possible future direction for audio inpainting.

\section{Conclusion}
In this paper, we built up the first benchmark for long audio inpainting, which could flourish the audio editing tasks. We propose deep spectrogram-based and waveform-based audio inpainting models and compare with baselines from related research. Our model is learning based and could recover long audio mask with superior performance quantitatively and qualitatively against baseline methods on both SC09 and ESC-50. We also explore the affect of different mask ratios and model architecture, and discuss possible future direction for long audio inpainting.
{\small
\bibliographystyle{aaai}
\bibliography{egbib}
}

\end{document}